\documentclass{webofc}
\usepackage[varg]{txfonts}   
\woctitle{?????????}
\begin{document}
\title{Fusion-fission probabilities, cross sections and structure notes of super-heavy nuclei}
%
%

\author{\firstname{Micha\l{}} \lastname{Kowal}\inst{1}\fnsep\thanks{\email{michal.kowal@ncbj.gov.pl}},
        \firstname{Tomasz} \lastname{Cap}\inst{1}\fnsep,
        \firstname{Piotr} \lastname{Jachimowicz}\inst{2}\fnsep,
             \firstname{Janusz} \lastname{Skalski}\inst{1}\fnsep,
             \firstname{Krystyna} \lastname{Siwek-Wilczy\'{n}ska}\inst{3} \and
             \firstname{Janusz} \lastname{Wilczy\'{n}ski}\inst{1}\fnsep\thanks{{deceased}}
}

\institute{National Centre for Nuclear Research, Ho\.za 69,
PL-00-681 Warsaw, Poland
\and
Institute of Physics,
University of Zielona G\'{o}ra, Szafrana 4a, 65516 Zielona
G\'{o}ra, Poland
\and
Institute of Experimental Physics, Faculty of Physics, University of Warsaw, Warsaw, Poland
          }

\abstract{%
Fusion - fission probabilities in the synthesis of heaviest elements are discussed in the context of the latest experimental reports.
Cross sections for superheavy nuclei are evaluated using "Fusion by Diffusion" (FBD) model.
Predictive power of this approach is shown for experimentally known Lv, Og isotopes and predictions given for Z=119,120. Ground state and saddle point properties as:
 masses, shell corrections, pairing energies and deformations necessary for cross section estimations are calculated systematically
within the multidimensional microscopic - macroscopic method based on the deformed Woods-Saxon single particle potential.
In the frame of FBD approach predictions for production of elements heavier than Z = 118 are not too optimistic.
For this reason, and because of high instability of superheavy nuclei, we comment on some structure effects, connected with the K-isomerism phenomenon which
could lead to a significant increase in the stability of these systems.
}

\maketitle
\section{Introduction}
\label{intro}

The uncharted region of the $Z$-$N$ plane can answer many questions of fundamental importance for science.
One of them is: "what can be the largest possible atomic number $Z$ of an atomic nucleus?".
The current answer was gained thanks to the
 tremendous advance in experimental studies achieved in recent years \cite{O1,Ut,GSI1,GSI2,TASCA} via fusion-evaporation reactions,
 once the element Z=118 has been synthesized \cite{O2}. Question if we can go still further waits for an answer as
 attempts to go beyond Z=118 hit two obstacles:
 (i) a difficulty or impossibility of making targets from Es and heavier
 actinides, (ii) reactions with heavier projectiles like: $^{50}$Ti, $^{54}$Cr,
  $^{58}$Fe, and $^{64}$Ni, did not produce any ERs up to now.
A characteristic feature of the fusion-evaporation reactions
leading to the synthesis of superheavy nuclei is enormous
hindrance of the fusion process itself. Consequently, the cross
sections for the synthesis of heaviest elements are measured in
picobarns or even femtobarns. It is believed that the hindrance
is caused by the highly dissipative dynamics of the fusing
system in its passage over the saddle point on the way through
the multidimensional potential energy surface from the initial
configuration of two touching nuclei into the configuration of
the compound nucleus.

The main issue therefore boils down to the problem of motion
in the potential surface with a local minimum (representing the meta-stable state) and the barrier protecting this minimum
which one should overcome to fuse two nuclei.
Taking into account the influence of the environment requires
addition of the damping force and the stochastic fluctuations.
In the case of so called: “fusion-by-diffusion” model (FBD) \cite{Swiatecki1}, the stochastic
process of shape fluctuation leads to the overcoming of the saddle point.
Fluctuating force, which is de facto caused by shape fluctuations, is assumed as a white noise
to obtain the solution of the Smoluchowski diffusion equation in the deformation space along the fission valley.
 Thus, basically we touch the famous Kramers problem \cite{Kramers}; the relaxation from the meta-stable state with the assumption
of the white noise and with damping proportional to the velocity.
Induced or spontaneous relaxation from the meta-stable state via fluctuating potential barriers occurs in many branches of physics:
phase transitions, bi-stability in quantum optic, spontaneous emission from excited states,
electron transport in semiconductors, kinetics of chemical reactions etc. \cite{Haken}.

The fundamental assumption which allows
to investigate the mechanism of the formation of superheavy
compound nuclei is the Bohr hypothesis, which states that the
synthesis of a compound system can be treated as a Markow-type
process (a stochastic process without any memory effect).
This implies that the exit channel is completely independent of the entrance
channel and both are independent of the intermediate stage of the reaction leading
to the compound nucleus. The Bohr hypothesis can be justified
mainly due to the different time scales of the particular stages.
According to this hypothesis the total probability for the
synthesis of a new superheavy element can be factorized into three
independent ingredients:
\begin{equation}
P_{tot}=P_{cap}\times P_{fus}\times P_{sur},
\label{Eq1}
\end{equation}
where $P_{cap}$ stands for the probability of overcoming the Coulomb
barrier, called the "capture process",  $ P_{fus} $ is the
formation probability that the nucleus, starting from the
touching configuration, will finish up with a compound nuclear
shape, and $P_{sur}$ - probability that the compound nucleus will
survive against fission. The cross sections for the synthesis of superheavy nuclei
are dramatically small because the fusion probability $ P_{fus} $
is hindered (in some reactions even by several orders of
magnitude) due to the fact that the saddle configuration of
the heaviest compound nuclei is much more compact than the
configuration of two colliding nuclei at sticking. Survival probabilities $P_{sur}$ are, in turn, extremely
sensitive to the height of the fission barrier, especially in case
of hot fusion reactions, because at each step of the deexcitation
cascade the competition between neutron emission and fission
strongly depends on the difference of energy thresholds for
these two decay modes.

Low probabilities of  for superheavy nuclei production and their high instability makes it worthwhile
to search for a long-lived exotic configurations. Obvious candidates are high-K isomers or ground-states,

\section{Model Fusion-by-diffusion with macroscopic - microscopic input}
\label{sec-1}

The FBD model \cite{Swiatecki1} in its extended
version\cite{Cap} serves to calculate cross sections for the synthesis of superheavy nuclei.
The extension consists in including the angular momentum dependence of all probabilities of successive stages in Eq. \ref{Eq1}.
In other words, for each angular momentum ($l$) the partial evaporation residue
cross section $\sigma_{ER}(l)$ for production of a given final nucleus in its ground state is factorized as the product of
the partial capture cross sections:
\begin{equation}
\sigma_{ER}(l) = \lambdabar^{2} \pi \sum_{l=0}^{\infty} (2l+1) P_{cap}(l) P_{fus}(l) P_{sur}(l),
\label{Eq1}
\end{equation}
where: $\lambdabar^{2} = \frac{\hbar^{2}}{2\mu E_{c.m.}}$ is the wavelength
($\mu$ is the reduced mass of the colliding system and  $E_{c.m.}$ is the energy in the center of mass).

It is assumed in the FBD model that after the contact of the two nuclei, a
neck between them grows rapidly at an approximately fixed
mass asymmetry and constant length of the system. This “neck
zip” is expected to carry the system towards the bottom of the
asymmetric fission valley. This is the “injection point,” from
where the system starts its climb uphill over the saddle in
the process of thermal fluctuations in the shape degrees of
freedom. Theoretical justification of the above picture of fast
zipping the neck was given in Ref. \cite{David}, where the later stage
of the stochastic climb uphill was described by solving the
two-dimensional Langevin equation. Theoretical location of
an effective injection point can be deduced from this model
\cite{David}. Also in a modified fusion-by-diffusion model \cite{Zu} the
location of the injection point was estimated theoretically.

Competition between neutron emission and fission can be resolved
in the spirit of the transition state theory with the help of the well known
formulas.
All details
regarding the calculations of the survival probability $P_{sur}$ can
be found in \cite{Cap}. In case of calculating multiple
evaporation (xn) channels a simplified algorithm \cite{Cap2}, avoiding the
necessity of using the Monte Carlo method, was applied.
The most sensitive point here is the knowledge of the fission barrier height.
The most important in fission barrier calculations is its reliability which, to large extent, is based on the model used for calculating energy of a nucleus as a function of
 deformation. Energy maps are necessary to appreciate fission barriers. In our case
 multidimensional energy landscapes are calculated within the
Microscopic - Macroscopic (MM) model based on the deformed Woods-Saxon potential \cite{WS}.
The Strutinski shell and pairing correction \cite{STRUT67}
 is taken for the microscopic part.
For the macroscopic part we used the Yukawa plus exponential model \cite{KN}
 with parameters specified in \cite{MUNPATSOB}. Thus, all parameter values
  are kept exactly the same as in all recent applications of the
 model to heavy and superheavy nuclei.
  Mononuclear shapes can be parameterized via spherical harmonics
 ${\rm Y}_{lm}(\vartheta ,\varphi)$ by the following equation of the nuclear surface:
    \begin{equation}
   R(\vartheta ,\varphi)= c(\{\beta\}) R_0
 \{ 1+ \sum _{\lambda=1}^{\infty}\sum _{\mu=-\lambda}^{+\lambda} \beta_{\lambda\mu}{\rm Y}_{\lambda\mu}\},
   \label{eq:radius}
\end{equation}
  where $c(\{\beta\})$ is the volume-fixing factor and $R_0$ is the radius of a spherical nucleus.
This parameterization has its limitations; certainly, it is not suitable for
 too elongated shapes.
 However, for moderately deformed saddle points in superheavy nuclei
 it excellently reproduces all shapes generated by other parametrizations,
  e.g. by \cite{Mol2000}, as we checked in numerous tests.
All details of calculation of ground state and saddle point properties as:
 masses, shell corrections, pairing energies and deformations can be  found in \cite{barKow,table}
However, in cited papers all tables are limited to even–even nuclei. This is why the
fission barrier heights for other nuclei have been calculated
separately by adding the quasiparticle energy:
$E_{q} =  \sqrt{(\varepsilon_{q}-\lambda)^2+\Delta^2}$ , where $\varepsilon_{q}$ is the energy of the odd nucleon
state $q$,  $\lambda$ is the Fermi energy and $\Delta$ is the pairing gap
energy. Calculations have been performed without blocking.

It should be emphasized that in the frame of this method the largest ground state shell effect ($\sim$ 9 MeV) is observed for the semi-magic nucleus $^{270}$Hs
 ($Z=108$, $N=162$) and the second minimum
 ($\sim$ 7MeV) of the shell correction is located around the nucleus $Z=114$ and $N=184$.
When superposed with weakly deformation-dependent macroscopic part,
 this component is largely responsible for the emergence of global minima in
 superheavy nuclei. Concerning the fission barriers, one obtains three areas with clearly raised barriers: around $N\approx$152,
 $N=$162 and $N\approx 180$, and the region of low barriers around $N=170$.
Finally, let us emphasize that our calculated first and second fission barriers in actinides are in a very good
  agreement with the experimental/empirical data: the root mean square deviation is 0.7 MeV \cite{IIbarriers} for the second, and 0.5 MeV \cite{barKow} for the first fission barriers.
In our opinion, such agreement justifies using our model for predictions in heavier nuclei.

\label{sec-2}

\section{Predictions for the synthesis cross sections of Z = 114–120}

\begin{figure}[h]
\centering
\includegraphics[width=7cm,clip]{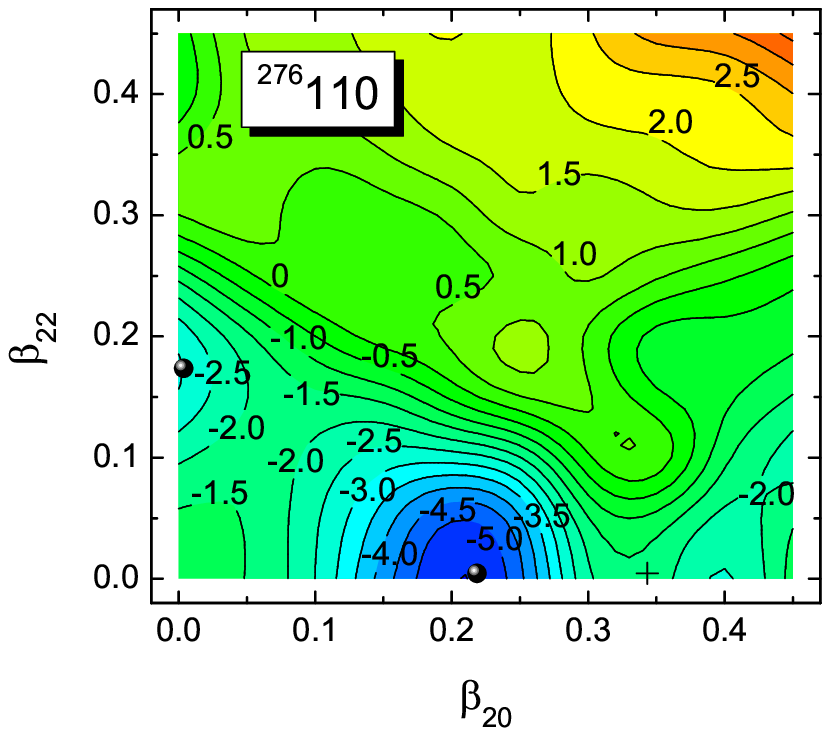}
\includegraphics[width=7cm,clip]{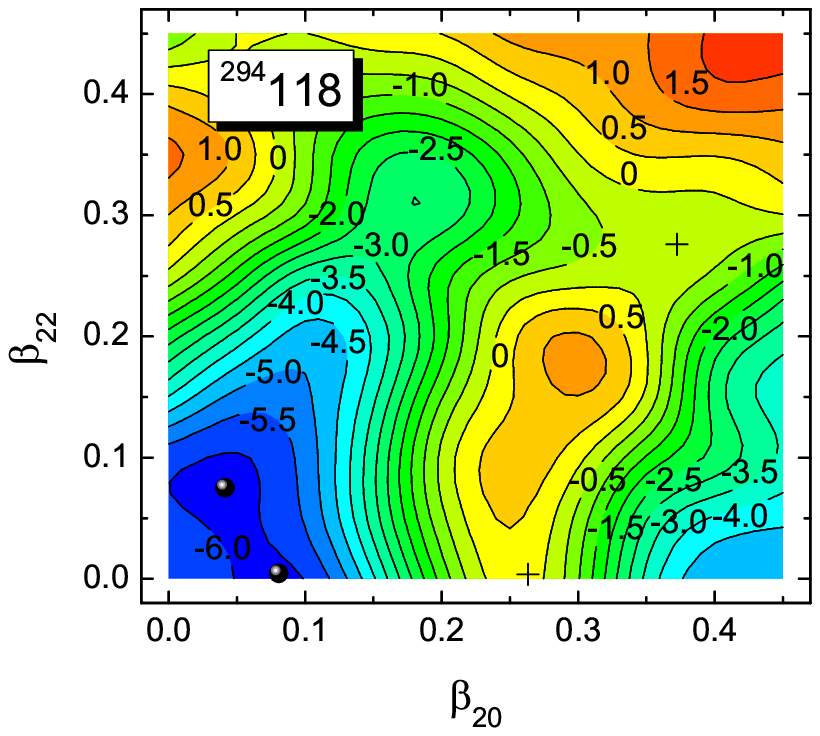}
\caption{Energy surfaces for two nuclei known experimentally.}
\label{fig1}       
\end{figure}

Since the inclusion of traxiality is absolutely necessary in the SH region, we start with showing
in Fig. \ref{fig1} two landscapes in the ($\beta_{20},\beta_{22}$) plane of experimentally known nuclei: $^{276}110$ and $^{294}118$.
In $^{276}110$ one can see two minima: prolate - the g.s. and a prolate superdeformed one, see left panel of Fig. \ref{fig1}. The map shows also the axial fission saddle.
The nucleus $Z=118$, $N=176$ is nearly spherical as can be seen in right panel of Fig. \ref{fig1}.
 There is also a secondary oblate superdeformed minimum. Competition between triaxial and axial saddles is clearly visible.

\begin{figure}[h]
\centering
\includegraphics[width=7cm,clip]{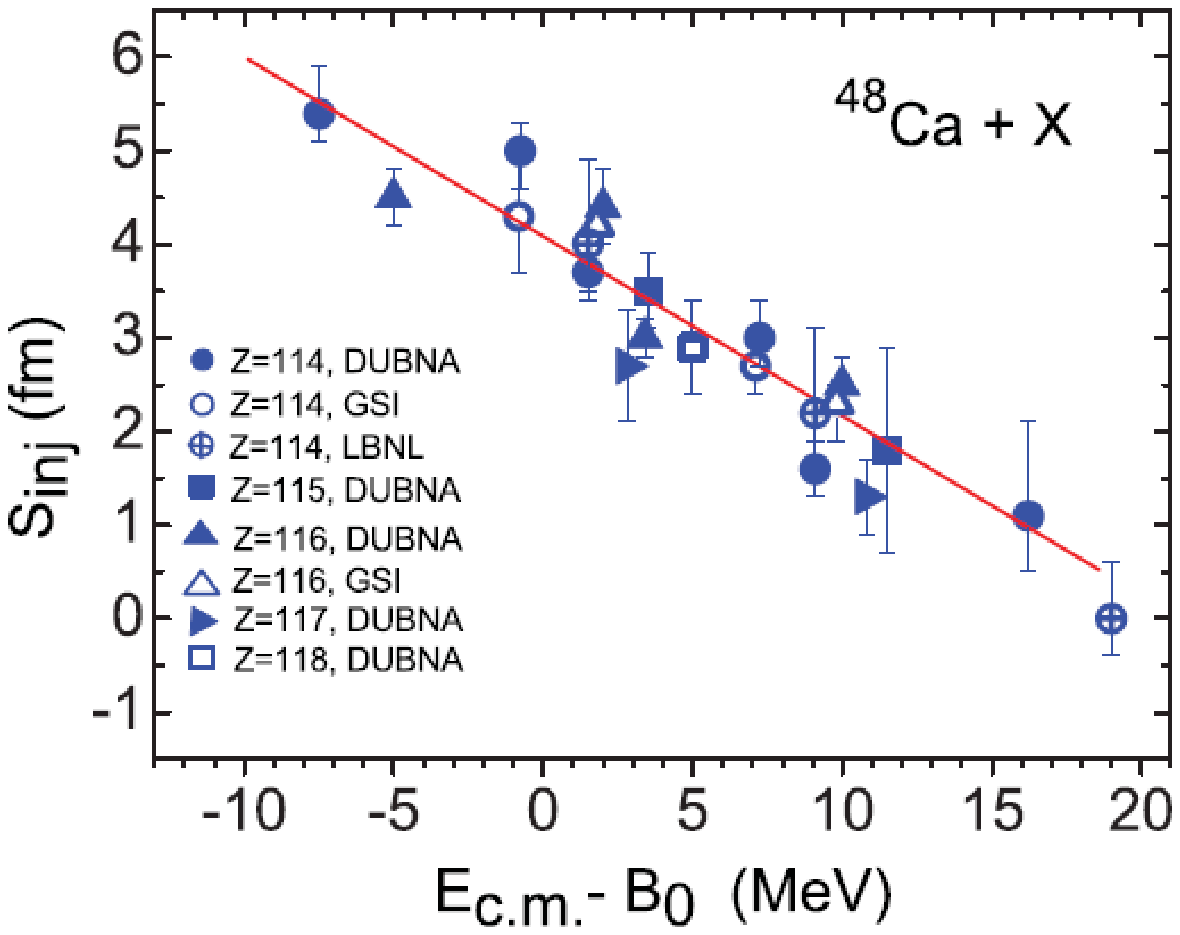}
\includegraphics[width=7cm,clip]{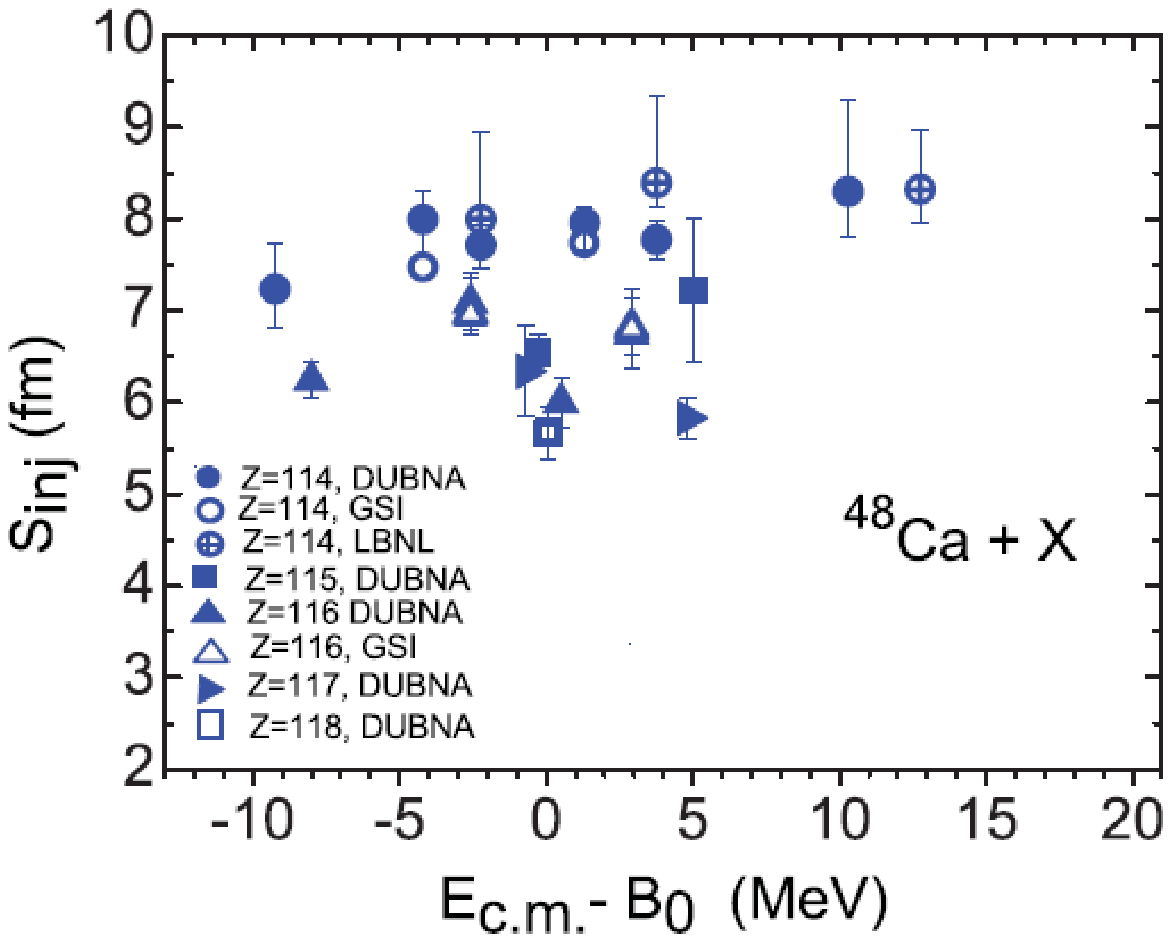}
\caption{Systematics of the injection-point distance $s_{inj}$ as a function of the kinetic energy excess determined using our macro-micro approach \cite{barKow,table} on the left and with P. M\"oller's compilation of nuclear data \cite{Mol1,Mol2} on the right.}
\label{fig2}       
\end{figure}

In the FBD model there is one fit parameter, namely $s_{inj}$, which is
defined as the excess of the total length of the combined
system over the length of the initial system (at the touching
configuration) when the neck-zip process brings the system to
the asymmetric fission valley. Based on the nuclear masses and fission barriers deduced from the energy maps (as shown above for $^{276}110$ and $^{294}118$) one can adjust
this parameter to the experimental synthesis cross section at the maximum of a given (xn) excitation function.
Details of this fit procedure with the specification of reaction data used can found in \cite{Wilczynska1}.
The compilation of the so-deduced $s_{inj}$ values is displayed in Fig. \ref{fig2}
as a function of the kinetic energy excess $E_{c.m.}$ - $B_{0}$ above the Coulomb barrier $B_{0}$ in the case of two sets of theoretical input data:
based on the deformed Woods-Saxon model \cite{barKow,table} described here (left panel) and assuming the fission-barrier heights \cite{Mol1} and
the ground-state masses \cite{Mol2} of P. M\"oller et al. (right panel).
Based on what we known from the dynamics of nucleus-nucleus
collisions one can expect that the injection distance
will increase with the decreasing energy:$E_{c.m.}$ - $B_{0}$.
It is exactly the case if we are using our data. Such a very good correlation between the $s_{inj}$ values and
the corresponding energies $E_{c.m.}$ - $B_{0}$ can be viewed as an
argument in favor of our macroscopic - microscopic results because such a striking correlation would be very
unlikely if the theoretical barrier heights were inconsistent with experimental values.
In contrast, the right panel of this figure
demonstrates the evident inconsistency of the set of $s_{inj}$ values obtained for the barriers of Ref. \cite{Mol1} in which the individual
points seem to be almost randomly scattered and do not show any correlation with energy.
\begin{figure}[h]
\centering
\includegraphics[width=7cm,clip]{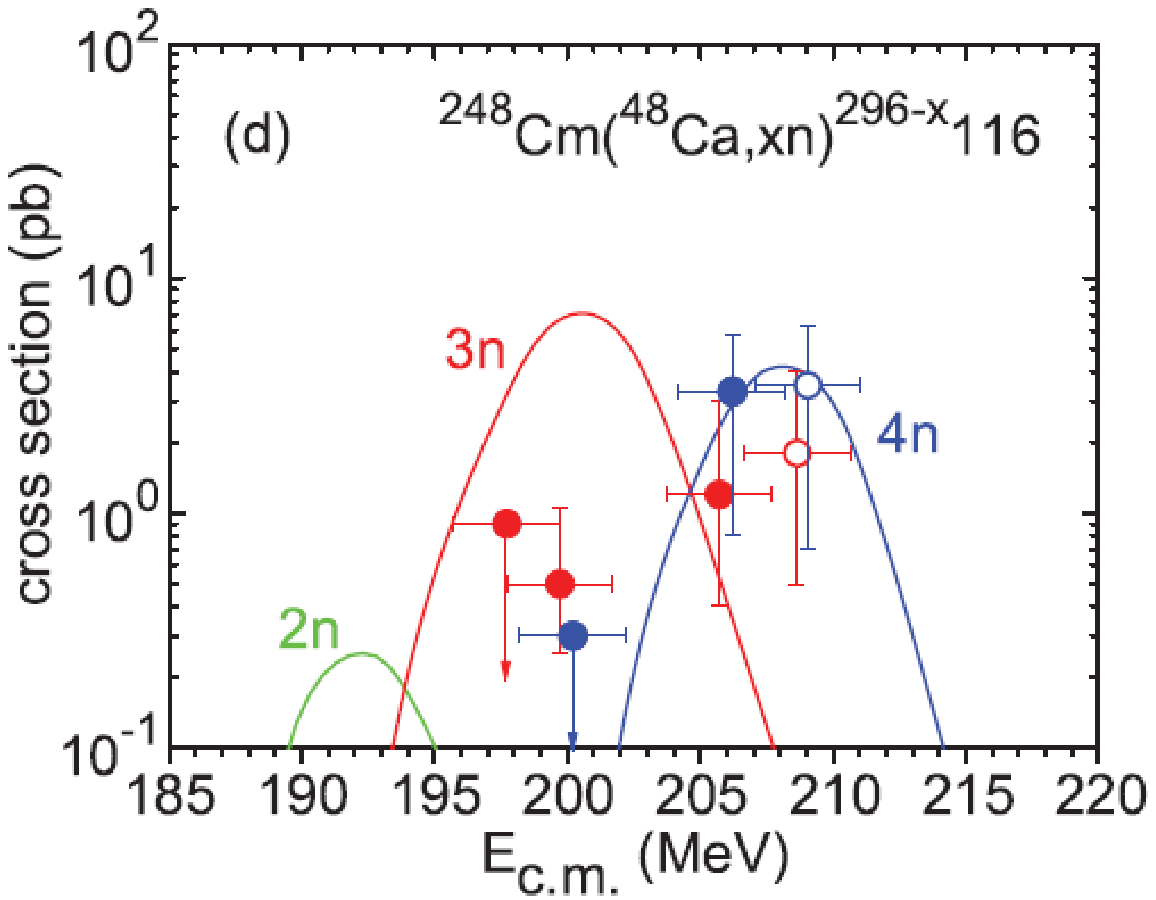}
\includegraphics[width=7cm,clip]{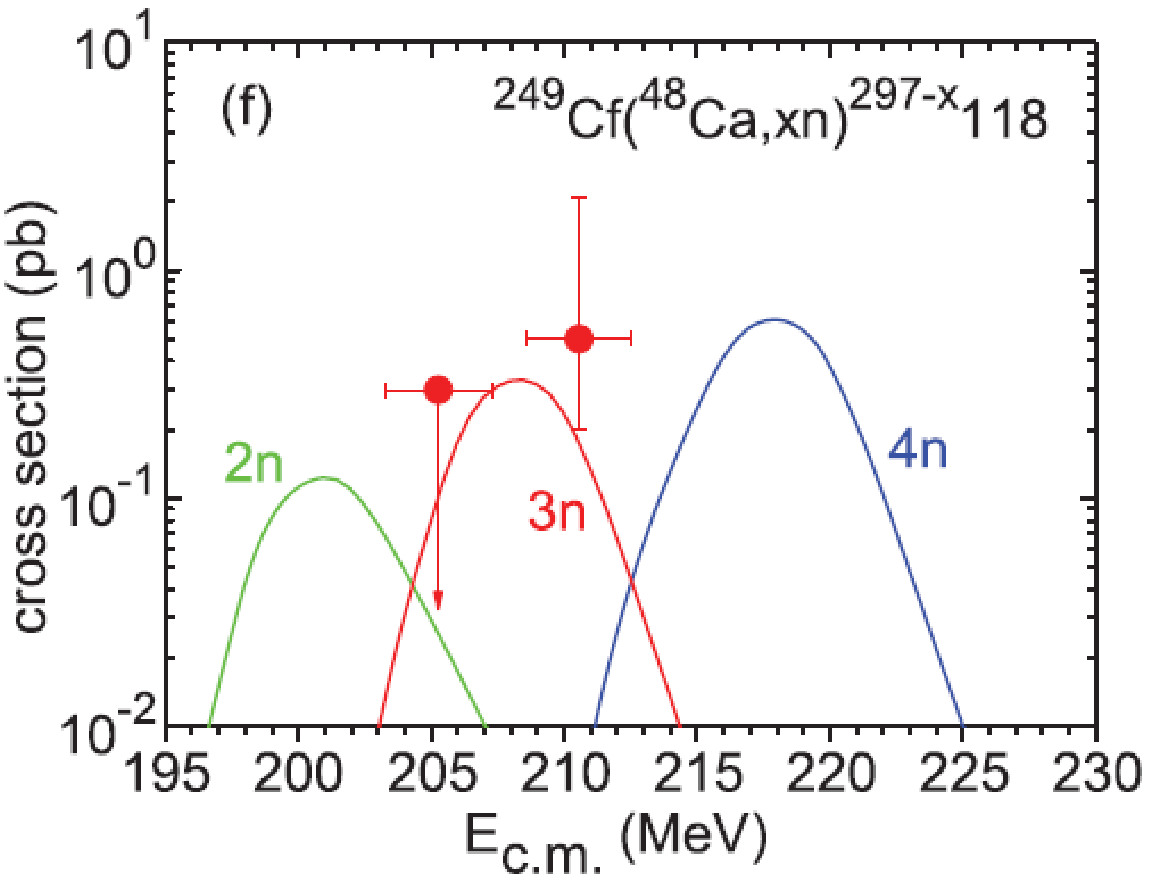}
\caption{Energy dependence of the cross section for synthesis of superheavy nuclei in hot fusion reactions. Full circles represent
data for 3n, 4n, and 5n reaction channels obtained in Dubna experiments for elements Z = 116 and Z=118. Open circles represents data taken from GSI.}
\label{fig3}       
\end{figure}

In \cite{Wilczynska1} we calculated the excitation functions for known superheavy elements starting from Flerovium up to Oganesson,
using results of our micro - macro model. In Fig. \ref{fig3} we show two examples, namely: $^{292}Lv$,$^{293}Lv$,$^{294}Lv$  and  $^{293}Og$,$^{294}Og$,$^{295}Og$.
Taking into account that the fission barriers could not be calculated more precisely than up to 0.5 MeV, the obtained agreement with experimental data can be considered satisfactory.

\begin{figure}[h]
\centering
\includegraphics[width=7cm,clip]{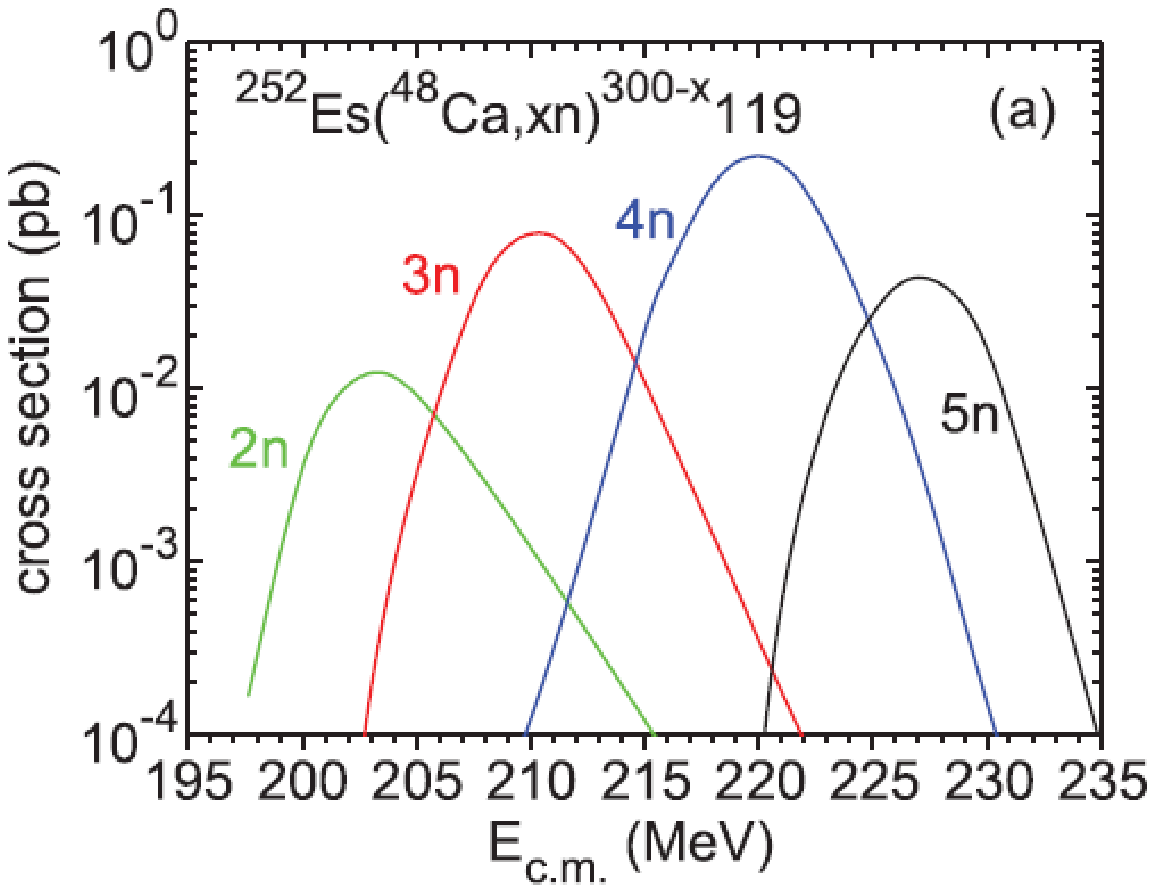}
\includegraphics[width=7cm,clip]{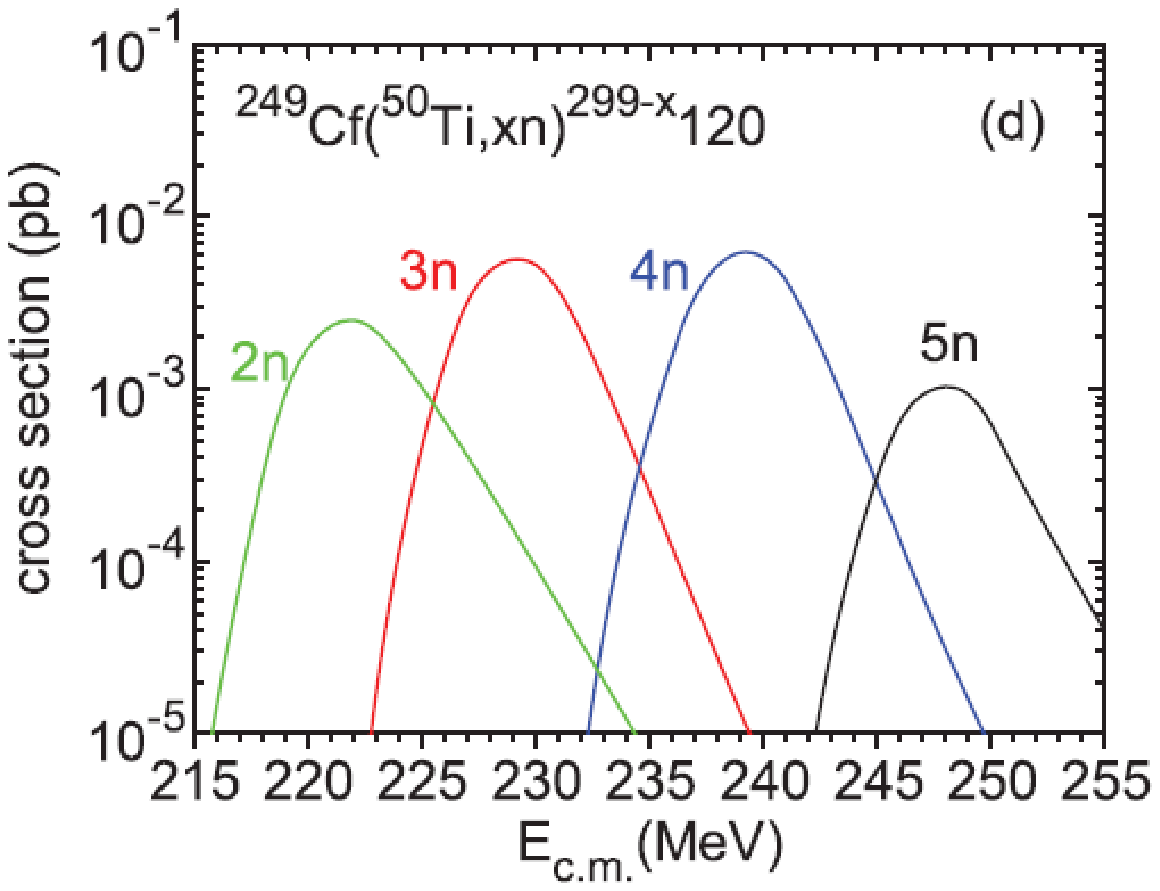}
\caption{Synthesis cross sections of yet undiscovered superheavy nuclei of Z = 119 and 120 predicted by using the
fusion-by-diffusion (FBD).}
\label{fig4}       
\end{figure}

Then we want to give predictions for currently unknown superheavy elements Z=119 and Z=120.
Particulary interesting is the nucleus $^{302}$120,
 as two unsuccessful attempts to produce it have already taken place in GSI,
 providing a cross-section limit of 560 fb \cite{120GSI1} or 90 fb in
 \cite{120GSI2}, and in Dubna \cite{120DUBNA}, providing the limit of 400 fb.
 The cross-section estimates \cite{Wilczynska3} do not support a possibility
 of an easy production of this SH isotope in the laboratory.
It is worth noting that with the barrier of the order of 10 MeV, as obtained in the
 frame of the self-consistent theory \cite{SKM}, producing superheavy Z=120 nuclei
  should not pose any difficulties.
We chose these two cases of fusion reactions for which our predictions are the most optimistic.
For Z=120, it turned out to be the reaction: $^{249}Cf(^{50}Ti,xn)^{299-x}120$. The corresponding excitation function is shown on the right side in Fig. \ref{fig4}.
The largest cross section about 6fb is expected for the 3n and 4n evaporation channels.
One very important remark is required here: carrying out this reaction would require a change
of the projectile from $^{48}Ca$  used successfully up to now to $^{50}Ti$ with which no hot SHE synthesis was completed successfully.
Generally, prospects for the synthesis of element Z = 120 are
considerably worse than those for Z = 119. To obtain Z=119, our calculations prefer the $^{252}Es(^{48}Ca,xn)^{300-x}119$
reaction, for which the synthesis cross section of about 0.2 pb in the 4n channel at $E_{c.m.} \simeq 220 $ MeV is expected. This case is shown on left panel in Fig. \ref{fig4}.

Closing this section we want to emphasise that the existing theoretical evaluations of the fission barriers differ
     significantly. Even the results of the two models based on the
   microscopic-macroscopic approach differ dramatically for some nuclei.
   Our calculations indicate, in contrast to the self-consistent mean-field
  studies, that fission barriers, still quite substantial for
    some $Z=118$ nuclei, become lower than 5.5 MeV for $Z=126$.
    Quite recently we noted a dramatic divergence in calculated fission barriers \cite{japan2015}.
    A brief review of recent progress in theoretical studies on fission barriers and fission half-lives of even–even superheavy nuclei can be found in \cite{nucl2015}.

\section{K-isomerism and possible stability enhancement }

A recent theoretical study \cite{BroSkal} of barriers within both the MM Woods-Saxon and Skyrme SLy6 Hartree-Fock plus BCS models
do not predict much chance to produce still (Z>126) heavier nuclei.
Moreover, it seems that two current methods of making superheavy elements in the laboratory: cold and hot fusion reactions, seem to reach their limits. 
On the other hand, not all superheavy (SH) isotopes Z < 118 have been produced yet. Therefore, while pondering upon possible new reactions leading towards the island of stability,
it may be worthwhile to search for long-lived exotic superheavy configurations.
Obvious candidates are high-K isomers, for which increased stability is expected 
due to some specific hindrance mechanisms. In our recent work \cite{Kisomers} we went even further: 
from systematic calculations it turned out that some nuclei can have the $\it ground state $  with a high-K quantum number.

\begin{figure}[h]
\centering
\includegraphics[width=7cm,clip]{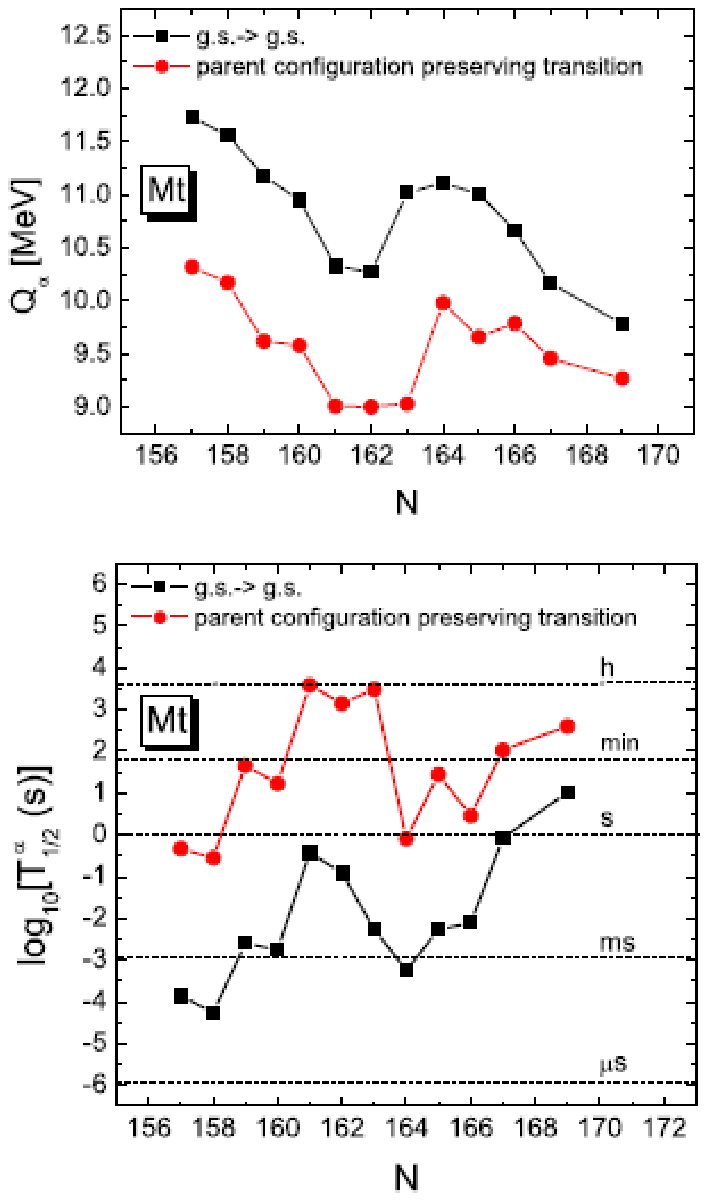}
\includegraphics[width=7cm,clip]{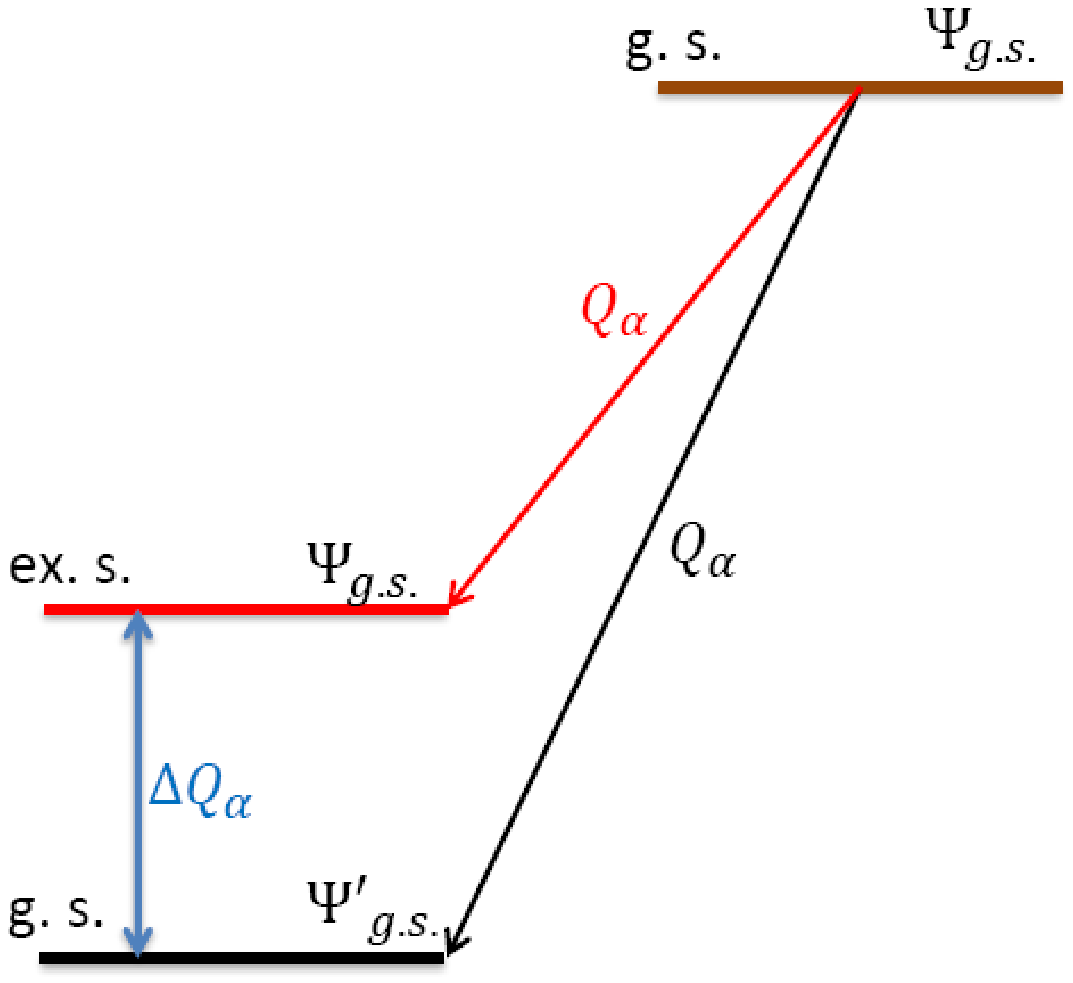}
\caption{$Q_{\alpha}$ values calculated with blocking following from the WS model
  and the increase in $\alpha-$ half-life induced by $\Delta Q_{\alpha}$
  \cite{Roy}, assuming a complete decay hindrance to
  different configurations (left panel). Schematic representation of appropriate transitions (right side).}
\label{fig5}       
\end{figure}

Structure of odd-odd nuclei is more complicated than that
of odd-A systems. If we disregard collective vibrations, the
ground state configuration is a result of coupling the unpaired
neutron and proton to a total angular momentum. The energy
ordering of coupled configurations is usually attributed to a
residual neutron-proton interaction. This is why, the investigation of odd nuclei requires a significant expansion of the used method.
This was done in Ref. \cite{JachKowSkal2014}.
An $\alpha$-decay hindrance of high-K configuration is expected when the same configuration in daughter has a sizable excitation.
If configurantion-changing transitions had been strictly forbidden, the hindrance would have been determined by this excitation
 $\Delta Q_{\alpha}$ which is schematically shown in Fig. \ref{fig5} (right panel). Also in Fig. \ref{fig5} (the left panel), a decrease in energy release,
 $\Delta Q_{\alpha}$, and $\log_{10} T_{\alpha}$ calculated
  according to \cite{Roy} for g.s.$\rightarrow$ g.s.  and
 configuration-preserving transitions are shown for
  various Meitnerium isotopes. The latter half-life would correspond
 to an absolute hindrance of configuration-changing decays.
 For our favorite case, $^{272}$ Mt, such hindrance reaches six orders of
 magnitude. This takes us from the life-time of milliseconds to hours.
  Let us stress that the whole argument is based on both the
 presence of deformed semi-magic shell Z=108 and N=162 and the
 position of high-$\Omega$ intruder orbitals just above that shell.

\section{Summary}

It may be that the heaviest element that can be created and identified will simply be limited by our synthesis techniques.
Predictions based on the FBD model are rather pessimistic, giving the cross sections at femtobarns level.
Synthesis of elements Z>118 will require a projectile heavier than $^{48}Ca$ (as fermium cannot be used as a target), and such reactions have thus so far proven fruitless.
Might it be possible, therefore, to obtain superheavy isotopes with a structure that slows down fission and alpha-decay?
Of particular interest would be half-lives of several seconds or even minutes, since this would make it possible to analyze the chemical properties of individual radioactive atoms.
The alpha-decay which dominates in the extreme superheavy region is delayed by differences in structure of parent and daughter configurations, resulting either from their very different deformations or a different angular momentum coupling of nucleons.

\section{ACKNOWLEDGMENT}
 M.K. and J.S. were co-financed by the National Science Centre under Contract No. UMO-2013/08/M/ST2/00257  (LEA COPIGAL). One of the authors
(P.J.) was cofinanced by Ministry of Science and Higher Education:
„Iuventus Plus” grant Nr IP2014 016073.

%
%

\end{document}